# High Fidelity Noise-Tolerant State Preparation of a Heisenberg spin-1/2 Hamiltonian for the Kagome Lattice on a 16 Qubit Quantum Computer

Wladimir Silva, Dept. of Computer Engineering, North Carolina State University, Raleigh NC.

This work describes a method to prepare the quantum state of the Heisenberg spin-1/2 Hamiltonian for the Kagome Lattice in an IBM 16 qubit quantum computer with a fidelity below 1% of the ground state computed via a classical Eigen-solver. Furthermore, this solution has a very high noise tolerance (or overall success rate above 98%). With industrious care taken to deal with the persistent noise inherent to current quantum computers; we show that our solution, when run, multiple times achieves a very high probability of success and high fidelity. We take this work a step further by including efficient scalability or the ability to run on any qubit size quantum computer. The platform used in this experiment is IBM's 16 qubit Gudalupe processor using the Variational Quantum Eigensolver (VQE).

## I. Introduction

For a long time, physicists have been modeling the ground and excited states of molecules in search for better materials and superconductors at high temperatures. A few years ago, supercomputers were the only choice to model even the simplest of molecules. That changed after 2018 when IBM opened up their new state of the art quantum computers to the scientific community. These new highly sophisticated quantum processors can be powerful tools to model these heavy elements.

### 1. Frustrated Systems

Kagome lattice is an arrangement of atoms in 2-D pattern commonly found in minerals. Figure 1 shows a basic unit cell of the lattice made up of a hexagonal star. Because of its geometry, this lattice is heavily studied in the field of condensed matter physics. When atoms in the lattice are anti-aligned by spin, the triangular shape creates an frustration (right side of figure 1), that is, the atom on the right doesn't know what to do as it is supposed to be anti-aligned with the one on the left and top at the same time.

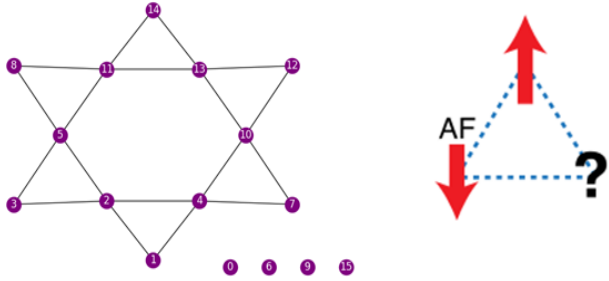

**Figure 1: On the left: Kagome lattice unit cell consisting of 12 vertices. On the right, an example of a frustrated system with atoms anti aligned by spin.**

This type of frustrated system is believed to be of great importance in the study of spin liquids and superconducting materials at high temperatures [1]. Modeling this type of frustrated system requires a tremendous amount of memory, and that's where a quantum computer can help.

### 2. Variational Quantum EigenSolver (VQE)

The VQE algorithm is made of three parts: An Ansatz or initial quantum circuit. This circuit is made of a set of rotation gates over the Y, Z axis of the Bloch Sphere parameterized by random angles. The Ansatz is coupled with a *cost function* whose task is to evaluate the Hamiltonian producing an energy value. This value is managed by a classical optimizer whose job is to minimize it, by updating the angles using sophisticated techniques such as: gradient descent, heuristics, and others. At the end, this sequence repeats for a number of cycles, ideally reaching the ground state. The energy values on each cycle are collected to produce a final plot of the energy minimization process. These values are compared with a classical eigensolver to estimate the fidelity (or accuracy) of the experiment. Noise can play a significant role in this process.

## II. State Preparation

Our experiment relies heavily on Qiskit's circuit library using VQE, parameterized for the Kagome lattice, an Ansatz (or initial state) and the spin ½ Hamiltonian. The experiment is submitted to IBMs hardware cloud using Qiskit Runtime.

### 1. Ansatz: EfficientSU2

EfficientSU2 is the Ansatz picked for this experiment (See figure 2). It uses a well-known heuristic pattern to prepare trial wave functions common in classification tasks for machine learning. It is made of rotations around the Y, Z axes of the Bloch Sphere, along with entanglements of neighboring qubits. The angles are initialized to random values.

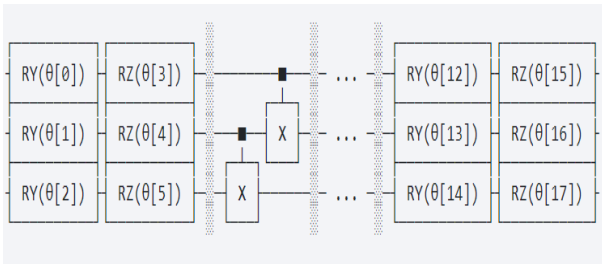

**Figure 2: EfficientSU2 Ansatz for 3 qubits with linear entanglements. Note that the RY-RZ-CX pattern repeats an arbitrary number of times (implementation specific).**

For the sake of simplicity, and to avoid noise accrual, a single number of repetitions are used.

### 2. Optimizer: NFT

This is the Nakanishi-Fujii-Todo algorithm [6]. In our simulations, when combined with EfficientSU2, it produces the best curve thus far: It descends quickly with a relatively low number of cycles (see figure 3).

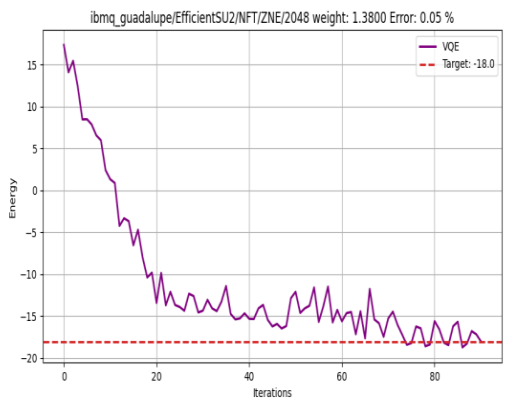

**Figure 3 NFT optimizer combined with EfficientSU2 descends quickly to the ground state with a low number of cycles. This experiment was run on IBM's Guadalupe processor and produced a very high fidelity of 0.05 below 1%.**

### 3. Hamiltonian

The Heisenberg spin ½ Hamiltonian (in figure 4). It is made up of interaction of the Pauli Matrices X, Y, Z over neighbor qubits. Note that, the sum (right side of figure 8) expands to 54 terms of observables (left side) each made of the product of a Uniform Interaction (initialized to the unit weight of the edge of the lattice) times a 16-tensor product of Identities (I) and Pauli Matrices mapped to the quantum processor qubit layout, producing a $2^{16}$ square density matrix. The Uniform interaction will play a critical role in error mitigation in our solution.

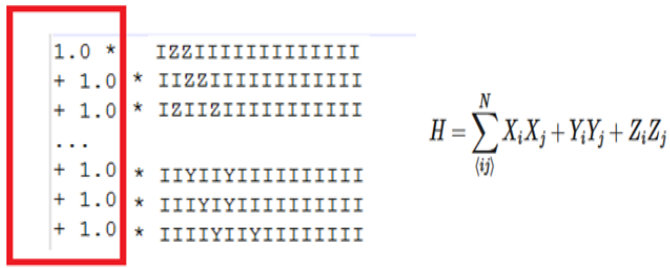

**Figure 4: The Heisenberg model designed to study critical points and phase transitions of magnetic systems where each site of a lattice represents a microscopic dipole with magnetic moment up or down.**

## III.   Experimental Results in Hardware

Initial experiments in the noiseless simulator were very encouraging. Results showed high fidelities, which we hoped will translate into hardware. We got lucky a few times on Hardware achieving high fidelity; nevertheless the overall results were different as shown by the failure rates in table 1. During the 4 months spent in this project, the noise accrued from two sources: entanglements (CX gates), and readouts (measurements). See figure 5.

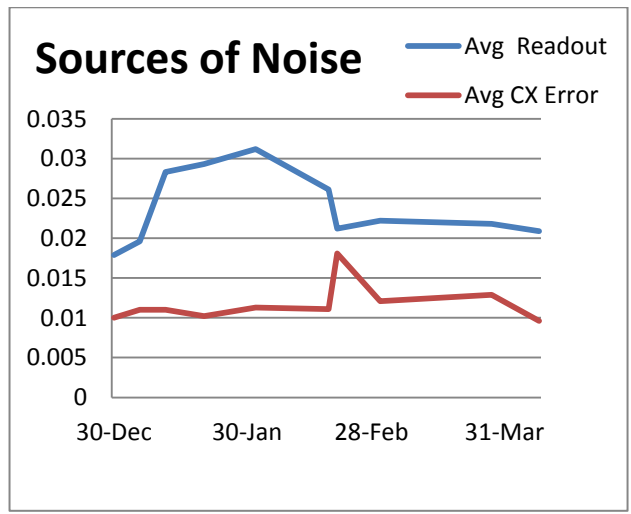

**Figure 5 shows the averages for the two main sources of noise for the 4 month period of this experiment. Readout errors crept up around twice the level of CX's.**

Average noise levels appear low: between 1-3% in CX and readout. However, the amounts accrue with a large number of qubits (12 readouts in our case) and 11 CX gates for a 1 repetition of the EfficientSU2 Ansatz.

1. **Results with Quantum Resilience**

Qiskit features a sophisticated quantum resilience architecture [7] which was included in our initial design. However, it made little difference if any. The levels of resilience are:

1. **Twirled readout error extinction (T-Rex):** It is a general and effective technique to reduce measurement noise by attaching extra measurement and calibration circuits for the estimation of error mitigated averages.
2. **Zero Noise Extrapolation (ZNE):** It works by first amplifying the noise in the circuit of the desired quantum state, obtaining measurements for several different levels of noise, and using those measurements to infer the noiseless result. *This resilience level is used in our work (see table 1).*
3. **Probabilistic error cancellation (PEC):** It samples for a collection of circuits that mimic a noise inverting channel to cancel out the noise in the desired computation similar to the way noise canceling headphones work.

Even though PEC provides the best theoretical resilience, by the time of this writing, its implementation is in alpha stage and its time complexity grows exponentially with the number of gates. In our experiments, PEC crashed after 6h of runtime.

**Table 1: Quantum resilience failure rates for 114 experiments on stage 1.**

| Method | Failure | Success |
| --- | --- | --- |
| **T-Rex** | 20 | 2 |
| **ZNE** | 84 | 7 |
| **PEC** | 1 | 0 |

2. **Initial Experimental Results without Mitigation**

We collected initial metrics from hardware to gain an insight on state fidelity. Besides the high error level, the noise tolerance was low: From a total of 114 experiments, 104 failed. A great deal of effort was put on collecting this data, with an average wait time of 4 days in the execution queue and a total of 170K jobs totaling 228h of quantum time.

| Metric | Description |
| --- | --- |
| **Total experiments** | 114 |
| **Failed** | 104 |
| **Avg # of VQE cycles/experiment** | 150 |
| **Total Jobs** | 17000 |
| **Avg execution time (h)** | 2 |
| **Total Quantum time (h)** | 228 |
| **Avg Queue wait time (days)** | 4 |

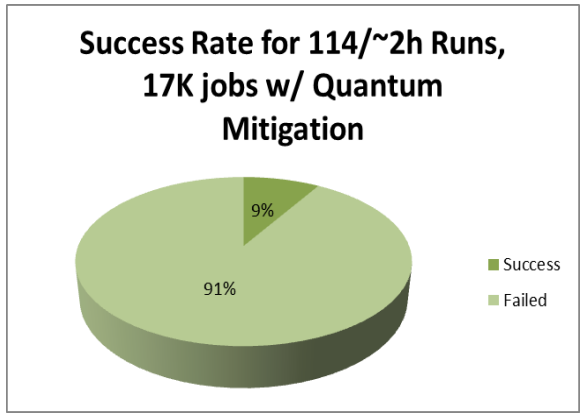

**Figure 6 Collected metrics of stage 1 of the solution with a low noise tolerance of 9%.**

### 3. Error Mitigation by Classical Means

To reduce the noise of a single experiment, and to build a solution that is noise tolerant over multiple runs, we considered the following options:

- Error correction codes: such as Steane[8] or Shor[9]. Unfortunately, this was out of the question due to the high number of ancilla qubits required to stabilize a single physical qubit. For example: Steane requires 7 ancilla per physical qubit, and Shor's requires 9. We only have a total of 16 qubits available so error codes are out.
- Classical post processing of data: via such methods as Fourier analysis, or other data massaging techniques. These methods fall outside our expertise and with little time to go they were a non-starter.

The only choice available was to try some sort of algorithmic post processing: Instead of massaging the data, massage the logic that consumes it.

### 4. Algorithmic Post-Processing Improvements

We decided to use Object Oriented Design techniques to wrangle the noise. A fundamental Object Oriented design principle states that objects should be immutable (this means, they cannot be changed after allocation). This has the benefit of preventing mistakes in highly concurrent environments. Thus, in our solution, arguments sent to the VQE (Ansatz, Hamiltonian, and Optimizer) are immutable by default. However, for this situation, the Hamiltonian was turned into a mutable object within the VQE so it can be corrected dynamically by applying four simple algorithmic rules.

1. If a point falls below the desired fidelity or error threshold of 1%, abort the process and return the collected data.
2. If a point falls below the target ground state by some delta (the distance between the point in the curve and the classic ground state), *dynamically decrease the uniform interaction* (UI) of the Hamiltonian, then continue. This has the effect of driving the curve upwards to the ground state.

3. If point falls above target by delta, do the inverse: *Increase the UI & continue. This drives the curve downwards* to the target ground state.
4. Recursion at the end: If at the end of the optimization cycle, the point ends above the target, recurse (repeat) from the last point: the optimization will resume towards the ground state. As a failsafe to prevent infinite recursions, a max 5 recursive calls is allowed. If at the end, the ground state is not reached, the experiment fails.

These four rules work in tandem with the probability of rule one being triggered at any stage of the process. This seemed like an unorthodox solution; nevertheless the results were pleasantly surprising (see Figure 7).

| Metric | Description |
|---|---|
| **Total experiments** | 76 |
| **Failed** | 1 |
| **Avg # of VQE cycles/experiment** | 100 |
| **Total Jobs** | 7600 |
| **Avg execution time (h)** | 1 |
| **Total Quantum time (h)** | 76 |
| **Avg Queue wait time (days)** | 6 |

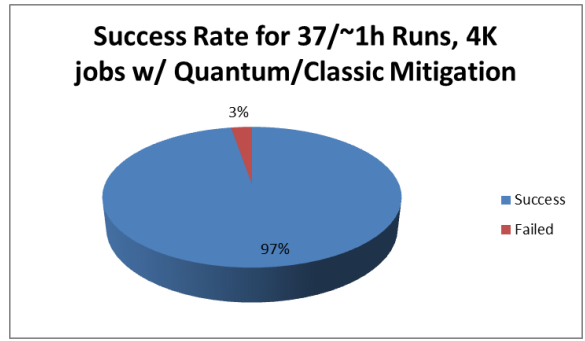

**Figure 7 Algorithmic post-processing achieved high noise tolerance (99%) and high fidelity (below 1%of relative error).**

## IV. Insights Gained and Discussion

A total of almost 5 months of work was spent in this experiment which gave us the following valuable insights on the feasibility of quantum computation for Ising model state preparation:

### 1. State Preparation was simple

Qiskit Runtime provides an extensive library of circuits and optimizers for state preparation, however there are some things to consider when picking an Ansatz-Optimizer combo which will determine the shape of the final curve:

1. **The combination of Ansatz, Optimizer:** It produces a different curve type. Consider figures 8, 9 and 10 which show different shape types from experimental results.

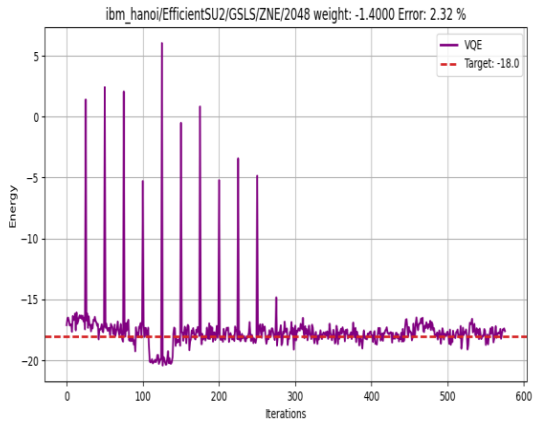

**Figure 8 shows the shape of the curve using the efficient SU(2) Ansatz[2] circuit consisting of two layers of single qubit SU(2) rotations and CX entanglements. It is coupled with the Gaussian-smoothed Line Search (GSLS)[3] optimizer using gradient approximation based on Gaussian-smoothed samples of a sphere. The experiment ran for about 4 hours in the 27 qubit Hanoi processor.**

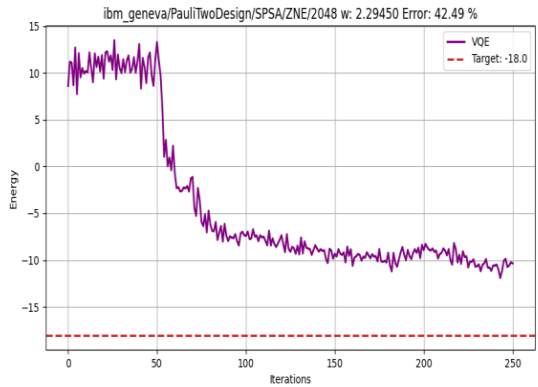

**Figure 9 shows an experiment in the 27 qubit Geneva running a PauliTwoDesign Ansatz, frequently used in quantum machine learning, along with a Simultaneous Perturbation Stochastic Approximation (SPSA) optimizer. This is a gradient descent method for large-scale population models [4]. It uses a relatively low number of cycles (250), the experiment ran for 3h.**

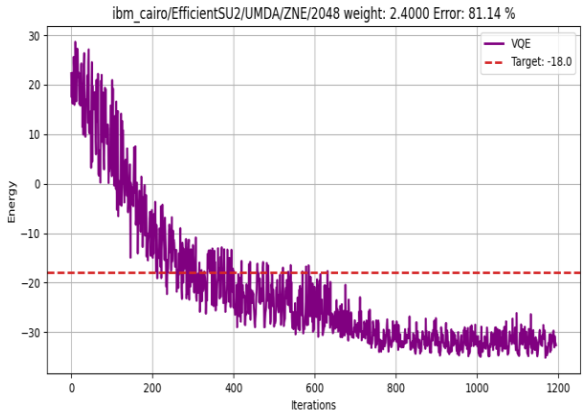

**Figure 10 shows a bizarre shape with a very large number of cycles run in the 27 qubit Cairo processor using the Univariate Marginal Distribution Algorithm (UMDA), a stochastic search from the family of the evolutionary algorithms. [5]. This experiment took an excessive 5+h of run time.**

2. **Number of cycles in the optimizer:** Some optimizers default to an excessive number of cycles. In figure 5, Univariate Marginal Distribution Algorithm (UMDA) defaults to 1200 cycles. This will balloon the execution time of the experiment, and worst of all, it will increase the odds of failure by memory exceptions or bugs in the server side. As a matter of fact, the experimental result from figure 5 was run in IBM's 27 qubit Cairo processor and took more than 5h to run.

### 2. Noise Mitigation took most of the Work

The supplemental materials show a pie chart of the work hours spent in this experiment over a period of 4 months. Almost 80% of the time was spent in noise mitigation as opposed to 1% in implementation. Noise management was the most difficult task even though the goal of the experiment was state preparation. Furthermore, noise affects not only a single experiment but the overall success rate over multiple executions.

### 3. Algorithmic Post Processing achieved high fidelity

Our results show that quantum resilience makes little difference in the final result fidelity. Furthermore other noise correction techniques such as error codes are impractical at this stage due to the excessive number of ancillary qubits required. The only way our experiment reached the desired fidelity was to use algorithmic post-processing: By dynamically updating the uniform interaction of the Hamiltonian, the energy curve zig-zags through the target ground state to reach a relative error below 1%.

## V. Conclusion

All in all, our solution for the Kagome lattice achieves high noise tolerance (it reaches a fidelity or relative error below 1% of the classical Eigen solver, 99% of the time). It does this using simple object oriented algorithmic post processing (as a matter of fact, the code that does the trick is less than 50 lines). When you run our experiment, it zigzags through the quantum noise with a high probability of success. This work was completed in a time span of 4 months with most of the time spent in error mitigation (see supplemental materials).

## VI.  Acknowledgements

We would like to thank IBM for opening these incredible machines to the academic and scientific communities for experimentation; as well as the IBM Quantum Awards – Open Science Prize 2022 for which this work was made.

## VII.  References

[1] Balents, L. Spin liquids in frustrated magnets. Nature 464, 199–208 (2010). https://doi.org/10.1038/nature08917

[2] Qiskit Circuit Library:
https://qiskit.org/documentation/stubs/qiskit.circuit.library.EfficientSU2.html

[3] GSLS from Circuit Library:
https://qiskit.org/documentation/stubs/qiskit.algorithms.optimizers.GSLS.html

[4] McClean et al., Barren plateaus in quantum neural network training landscapes. arXiv:1803.11173

[5] Qiskit UMDA circuit library
https://qiskit.org/documentation/stubs/qiskit.algorithms.optimizers.UMDA.html

[6] NFT optimizer available at https://arxiv.org/abs/1903.12166

[7] Error suppression and error mitigation with Qiskit Runtime.
https://qiskit.org/documentation/partners/qiskit_ibm_runtime/tutorials/Error-Suppression-and-Error-Mitigation.html

[8] Steane, Andrew (1996). "Multiple-Particle Interference and Quantum Error Correction". Proc. R. Soc. Lond. A. 452 (1954): 2551–2577. arXiv:quant-ph/9601029. Bibcode:1996RSPSA.452.2551S. doi:10.1098/rspa.1996.0136. S2CID 8246615.

[9] W.Shor, Peter (1995). "Scheme for reducing decoherence in quantum computer memory". Physical Review A. 52 (4): R2493–R2496. Bibcode:1995PhRvA..52.2493S. doi:10.1103/PhysRevA.52.R2493. PMID 9912632

# VIII. Supplemental Materials

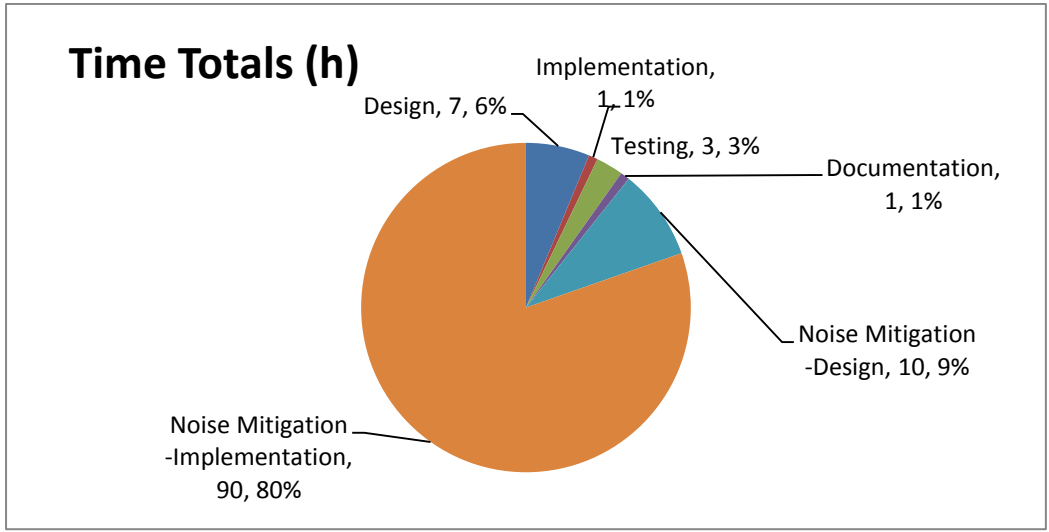

**Figure 11 Break down of the time (in hours) spent on this project over a period of 4 months.**

This is a sample of the experimental data collected over a period of 4 months. The choice of Ansatz and optimizer drives the shape of the curve. Note that the number of cycles is important too, as excessive amounts may run for a long time and may crash from bugs or memory exceptions on the server side.

Table 2 Results from miscellaneous Ansatz, Optimizer combos in Qiskit's circuit library.

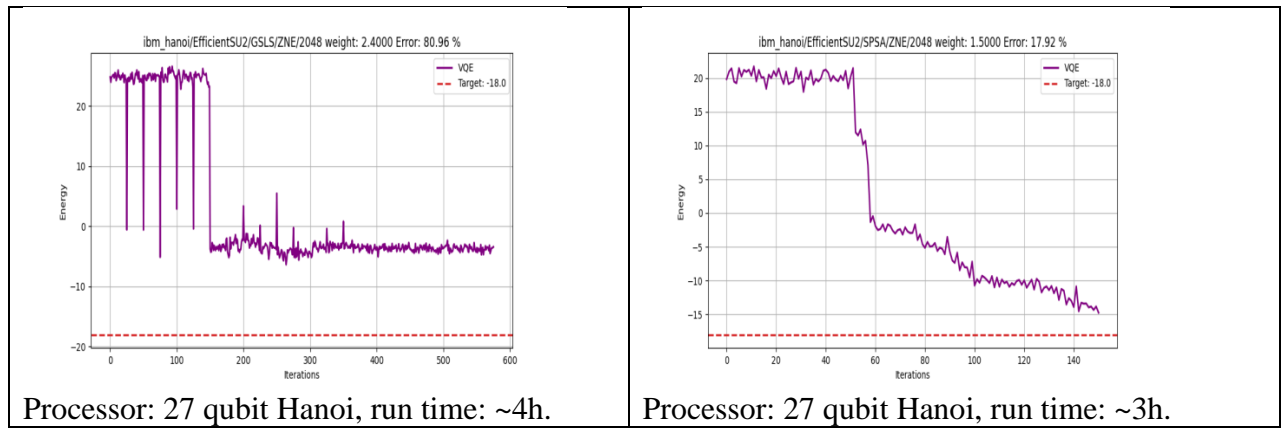

| Processor: 27 qubit Hanoi, run time: ~4h. | Processor: 27 qubit Hanoi, run time: ~3h. |

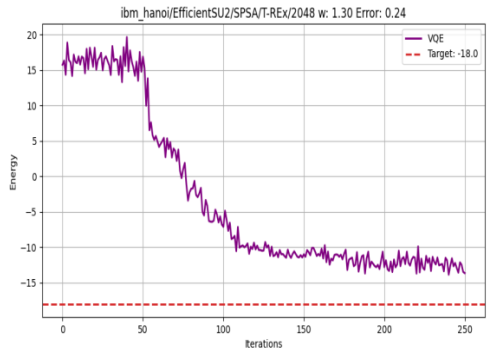

Processor: 27 qubit Hanoi, run time: ~2.5h.

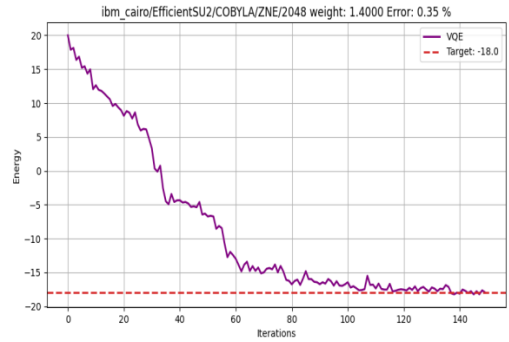

Processor: 27 qubit Cairo, run time: ~2h

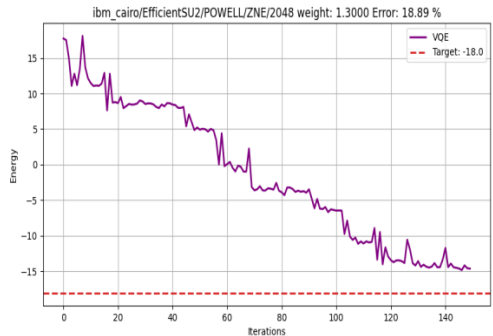

Processor: 27 qubit Cairo, run time: ~1.5h

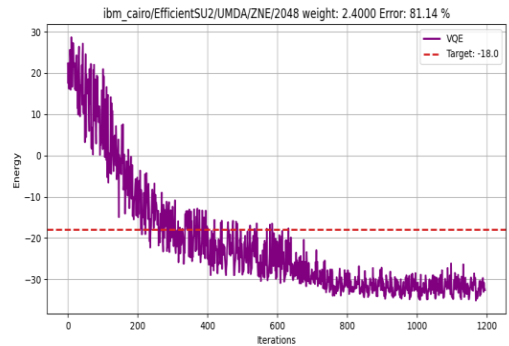

Processor: 27 qubit Cairo, run time: ~6h

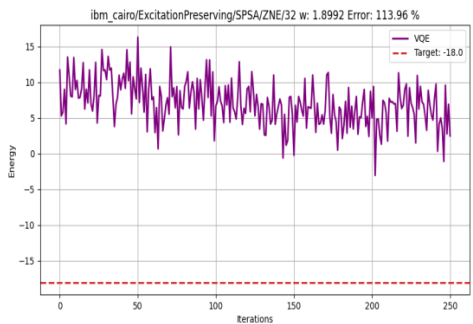

Processor: 27 qubit Cairo, run time: ~3h

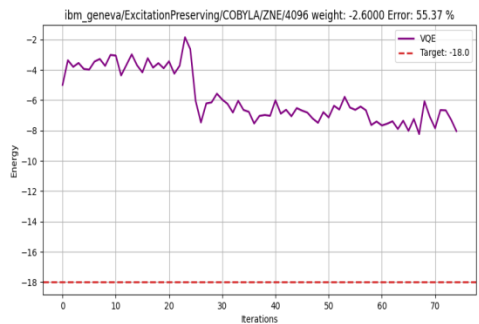

Processor: 27 qubit Geneva, run time: ~1h

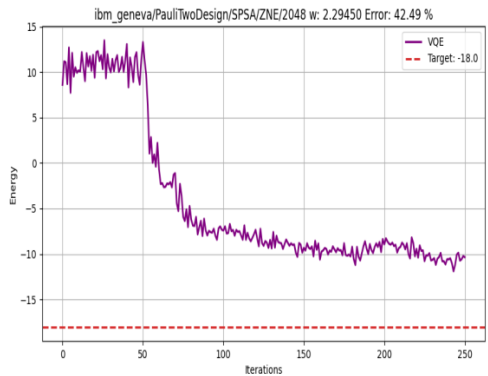

Processor: 27 qubit Geneva, run time: ~2h

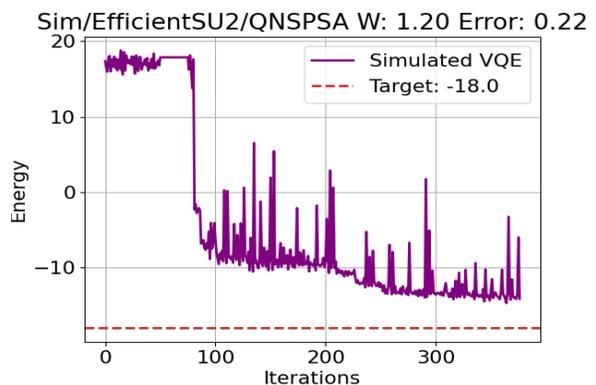

A simulator curve, too fast to compare with HW.

Legend - Optimizer

- GSLS: Gaussian-smoothed Line Search.
- SPSA: Simultaneous Perturbation Stochastic Approximation.
- COBYLA: Constrained Optimization by Linear Approximation.
- POWELL: The Powell algorithm performs unconstrained optimization.
- UMDA: Univariate Marginal Distribution Algorithm.
- QNSPSA: Quantum Natural SPSA.

Legend – Resilience

- ZNE: Zero Noise Extrapolation.
- T-Rex: Twirled readout error extinction.